# Oscillation Conditions of the Vircator Klystron with External Delayed Feedback: A Computer Simulation

**V. G. Anfinogentov and A. E. Hramov**

**Abstract**—A new version of the virtual-cathode oscillator, vircator klystron, is suggested and studied numerically. It essentially includes adjustable external feedback. The non-steady-state nonlinear dynamics of the oscillator is analyzed in relation to the beam current and feedback parameters. Physical processes in the electron stream during transitions between oscillation conditions are considered.

## INTRODUCTION

High-power microwave generation by virtual-cathode (VC) oscillators, called vircators, has been under active experimental and numerical investigation (see [1–3] and references therein). Special attention is payed to the VC devices that include different types of external or internal feedback [4–7]. Vircators with external delayed feedback are referred to as virthodes [4, 5]. They allow one to increase the efficiency and to control the spectrum and power of the output signal. The latter can be done by adjusting feedback parameters, e.g., the delay [4, 8, 9]. Also promising are internal-feedback vircators with an input resonator, called vircator klystrons [10, 11]. They could offer high efficiency and an almost sinusoidal output waveform.

This paper considers a vircator klystron provided with *external* delayed feedback, the input resonator being tuned to the VC frequency. Thus, the oscillator has two feedback loops with differing delays. Besides the external loop with delay $d$, a feedback is effected by the reflection of the electron stream from the VC to the input resonator. Both delays are less than $T_{VC}$, the characteristic period of VC oscillations. The presence of two feedback loops enables one to selectively control structures that are formed in the electron stream and determine the dynamics of the VC (see, e.g., [12–16]).

We take a numerical approach whereby electron motion equations and Maxwell's equations are solved in a self-consistent manner. On this basis, we explore the oscillation conditions of the device and physical processes in the electron stream.

## FORMULATION OF THE PROBLEM

Let us consider the following model of the vircator klystron with external delayed feedback. The oscillator contains a working chamber formed by a cylindrical waveguide of length $L$ and radius $R$. The waveguide consists of a hollow and a coaxial section extending over the intervals $0 < z < l$ and $l < z < L$, respectively. The coaxial section is filled with a homogeneous conducting medium of conductivity $\sigma$, which simulates the coaxial output used to extract the generated energy. The values of $L - l$ and $\sigma$ are selected to provide a power reflection coefficient below 10%. Specifically, we set $L = 16$ cm, $l = 10$ cm, and $R = 3$ cm. Let a single-velocity axially symmetric hollow electron beam of radius $r_b = 2$ cm, current $I$, and energy 560 keV be injected into the working chamber through a narrow coaxial input resonator. The beam current is assumed to be above the space-charge-limited current for the configuration at hand, so that a VC is formed in the beam. Electrons reflected from the VC travel through the resonator to be collected by an input grid. Accordingly, they are not involved in the stream oscillation occurring in the cathode–VC space. (We follow the so-called reditron model described in [17].) The resonance frequency of the input resonator is selected so that it is approximately equal to that of the VC; it is on the order of 8 GHz. The signal from the output part of the device comes to an input modulator via an external feedback circuit with delay $d$.

A strong magnetic field is applied along the system axis, so that the electron motion can be considered one-dimensional. For a given $z$, the cross section of the electron stream is viewed as a set of concentric charged annuli in order to allow for the nonuniformity of current-density distribution over the waveguide cross section. This technique is frequently used when dealing with klystron oscillators or amplifiers (see [18] and references therein). The dynamics of the $i$th annulus obeys the relativistic motion equations [19]

$$\frac{dp_{i,N}}{dt} = \frac{e}{m\gamma}E_z(r,z), \quad \frac{dz_{i,N}}{dt} = v_{i,N}, \qquad (1)$$
$$p = v/\sqrt{1-(v/c)^2},$$

where $e$, $m$, $z$, $v$, and $\gamma$ are respectively the charge, mass, coordinate, velocity, and relativistic factor of the annulus. Its radius is $r_N = r_0 + N \cdot \Delta r/N_\Sigma$, where $r_0$ is the inner beam radius, $\Delta r$ is the beam thickness, $N_\Sigma$ is the total number of annuli in the cross section, and $N$ ranges from 1 to $N_\Sigma$. To find a self-consistent field $E$, we numerically integrated Maxwell's equation system

$$\mathrm{curl}\vec{E} = -\frac{1}{c}\frac{\partial\vec{H}}{\partial t}, \quad \mathrm{curl}\vec{H} = \frac{1}{c}\frac{\partial\vec{E}}{\partial t} + \frac{4\pi}{c}\vec{j} \qquad (2)$$

by the finite-difference method using a two-dimensional spatial mesh [19, 20]. Motion equations (1) were solved by the leapfrog method [19].

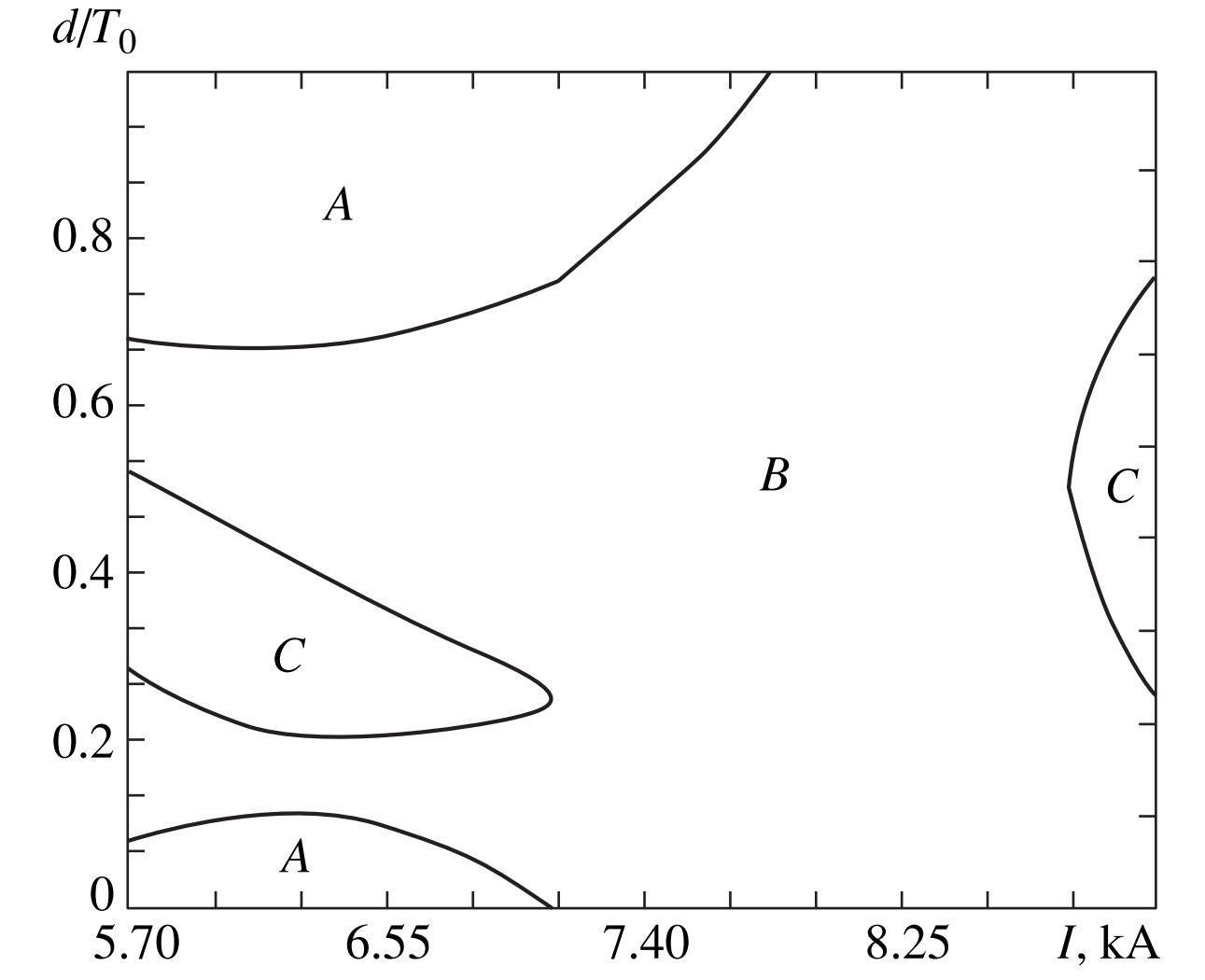

**Fig. 1.** Regions of oscillation conditions.

## TYPICAL OSCILLATION CONDITIONS

Let us explore oscillation conditions in relation to key control parameters, namely, the beam current $I$ and the external-feedback delay $d$. In the computer simulation, $I$ was varied over the range 5–10 kA and $d$ was varied from 0 to $T_0$, where $T_0$ is the natural period of the input resonator. The power spectra of oscillations generated in different oscillation conditions were computed from waveforms of the longitudinal electric-field component in the output (coaxial) waveguide. For the $I$–$dT_0$-plane, regions of main oscillation conditions are approximately outlined in Fig. 1. Notice that three oscillation conditions are possible, which respectively correspond to regions $A$, $B$, and $C$.

Region $A$ refers to almost sinusoidal oscillations, with the spectrum having a prominent principal component and a number of low ones. Figure 2a shows a power spectrum typical of this regime, for $I$ = 6.3 kA and $d/T_0$ = 1. There are two base frequencies, whose ratio is an irrational number. The prominent spectral component is located at one of them. The power generated at the second base frequency is lower by more than 20 dB. The remaining components are harmonic and combination frequencies. The spectrum includes a rippled noise background, which rapidly decreases with increasing frequency. Its magnitude is about 40 dB at 10 GHz.

Region $B$ corresponds to similar oscillations, but the power at the second base frequency is somewhat higher, –15 dB. The level of the noise background is higher as well.

Region $C$ is related to far more complicated oscillations. The spectrum now exhibits two almost equal components at two incommensurable base frequencies, with the noise background being stronger than that for region $B$. Figure 2b exemplifies this in the case $I$ = 6.3 kA and $d/T_0$ = 0.25. Note that biperiodic operation is fairly common among vircators (see, e.g., [2, 21]).

Thus, the vircator klystron with external delayed feedback, the input resonator being tuned to the natural frequency of the VC, has three main operating conditions: near sinusoidal oscillation (region $A$), weak quasi-periodicity with two base frequencies (region $B$),

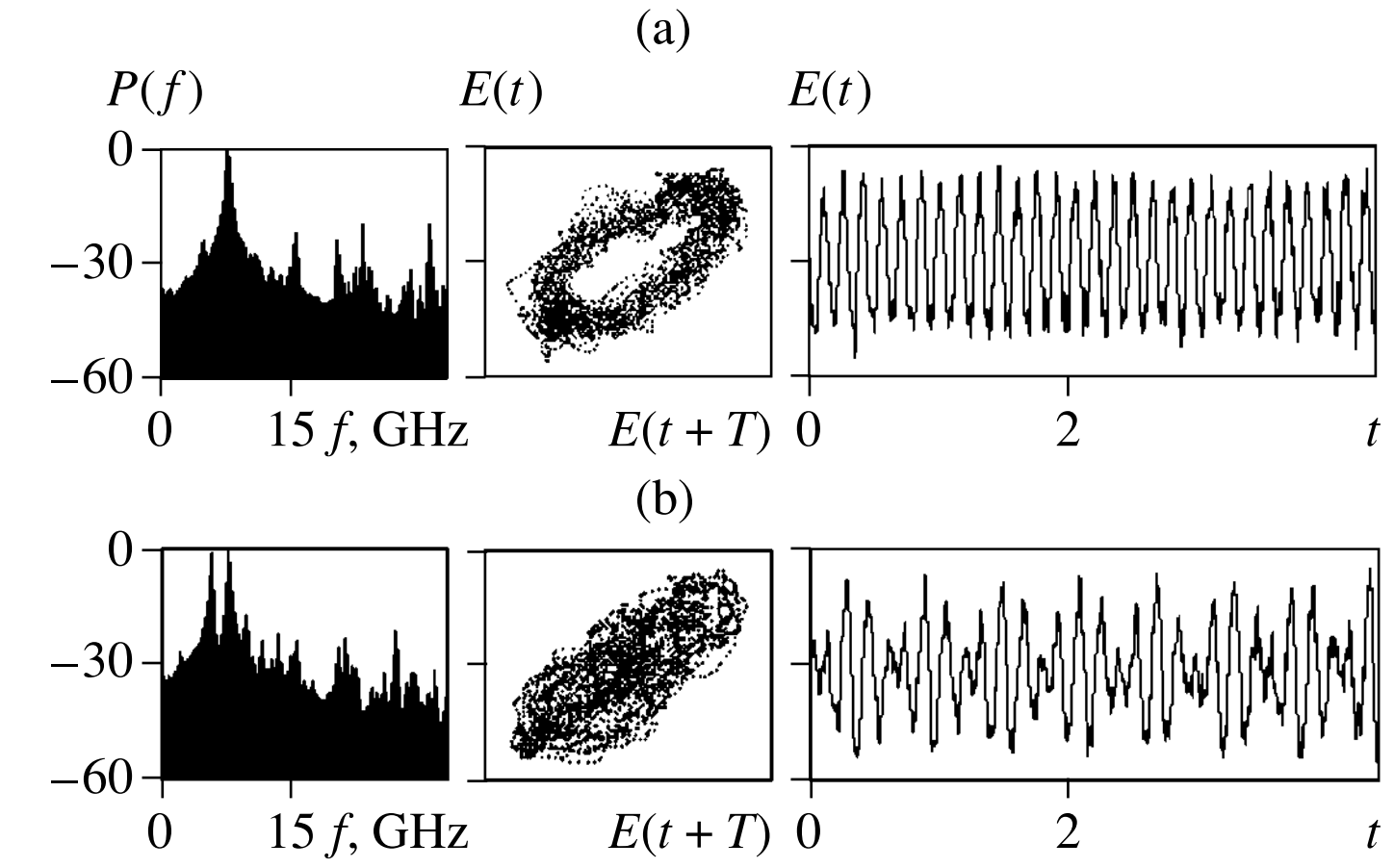

**Fig. 2.** Output power spectra for regions (a) $A$ and (b) $C$ in Fig. 1.

and chaos in the form of fully developed biperiodic oscillation (region *C*).

## PHYSICAL PROCESSES IN THE ELECTRON STREAM

Let us analyze the physical processes that attend transitions between oscillation conditions as the external-feedback delay $d$ is varied. Our conclusions will be based on trajectory plots in the $t$–$z$-plane and the distributions of reflected electrons in the lifetime in the interaction space, $F(\tau)$, the electrons being reflected from the VC back to the injection plane. The trajectory plots will refer to different layers of the stream, and the distributions, to different operating conditions.

Figure 3 shows trajectory plots of charged particles for three layers differing in radius. The motion is computed for one oscillation period. Figure 3a is drawn for the innermost layer. It illustrates how a VC typically develops in a vircator. It is seen that secondary structures formed in the transit stream decelerate in the time-dependent field of a developing VC (see, e.g., [13, 14]). The charged particles cluster into electron bunches, which partly reflect the stream back to the injection plane. (In the trajectory plot, the bunches are represented by the areas where electron trajectories converge.) The reflected portion of the stream disturbs the development of the VC in subsequent oscillation periods, thus complicating system dynamics [13–16]. In the middle layer (Fig. 3b), the formation of a VC and related phenomena are less noticeable. In the outermost layer (Fig. 3c), a VC exists for a short time and so do reflected particles; consequently, no kinematic instability is revealed in the zero-Coulomb-repulsion approximation. Also note that respective VCs come into being at different instants, starting with the innermost layer, as indicated by reflected electrons. Accordingly, the lifetime of the blocking potential barrier is maximal in the innermost layer and minimal in the outermost one. The beam thus develops a space-charge oscillation in the transverse direction.

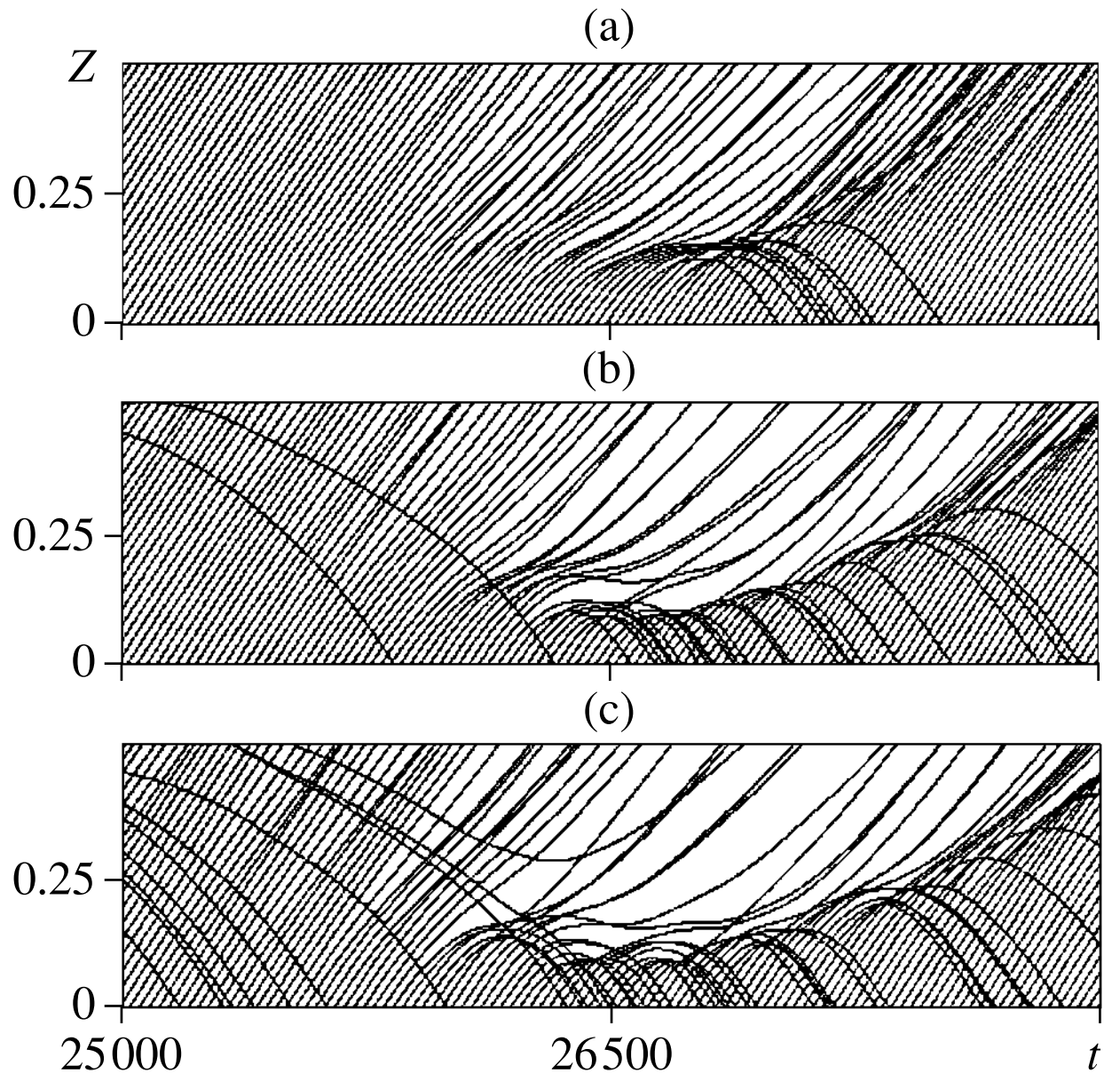

**Fig. 3.** Trajectory plots for three layers of an electron stream with $I$ = 6.3 kA. Panels (a), (b), and (c) refer to the outermost, the middle, and the innermost layer, respectively.

The interaction between structures in the layers leads to the onset of chaos in VC dynamics, so that the operation band widens. Furthermore, the formation of additional structures is strongly influenced by the feedback signal, as evidenced by the fact that the oscillation becomes more complicated if $d/T_0 \approx 0.3$–$0.5$. This is due to a nonzero phase angle that arises between the feedback signal and the field oscillation in the input resonator, the latter governing the premodulation of the stream.

To find out why the second independent base frequency appears in the spectrum, we examined the distributions of charged particles in their lifetime in the drift space, $F(\tau)$. Figures 4a and 4b show graphs of $F(\tau)$ for region *A*, at $I$ = 6.9 kA and $d/T_0 = 1$, and region *C*, at $I$ = 6.9 kA and $d/T_0 = 0.25$, respectively. They refer to the middle layer.

Let us consider the case where lifetime $\tau$ is not very long. For region *A*, the distribution has only one peak (see the left side of Fig. 4a). This indicates that there is a single spatial structure, namely, a VC, with a principal spectral component representing the oscillation of the VC as a whole.

By contrast, there are two peaks in the distribution corresponding to region *C* (see the left side of Fig. 4b). One of them is located in the same place as the above-mentioned peak, whereas the other (a highly blurred one) is typical of this regime only. It follows that two electron structures reflecting charged particles are formed in the electron stream (see, e.g., [5, 15]). The respective characteristic lifetimes of particles in the structures are in the same ratio as the base frequencies. Since $d/T_0 = 0.25$, we have $d \approx 0.5\langle T_e \rangle$, where $\langle T_e \rangle$ is the average transit time of an electron reflected from the VC. Accordingly, the velocity modulation of the stream is maximal at the instants when a VC is created in the stream and when the space-charge density $\rho$ near the injection plane is minimal. In this case, more structures are formed as a result of both the modulation of the stream by a strong feedback signal and zero-Coulomb-repulsion bunching. The space charge cannot impede the growth of the structures to an appreciable degree, because $\rho$ is small. The charges of the electron structures (dense bunches) are large enough to partly reflect the electron stream (see, e.g., [13–16]). This promotes instability in the stream, so that the oscillation becomes strongly chaotic, which we observe in region *C*.

When $d/T_0 \sim 1$ (Fig. 4a), we can see a low local peak at sufficiently large $\tau$ as well as the principal one located in the region of small lifetimes. The lifetime at which this low peak is located is important in that the dynamics of corresponding particles is responsible for the generation of the second (low) base frequency in the power spectrum. This time scale of the stream dynamics cannot be noticed with ease in the distributions or power spectra. Nevertheless, as *d* is decreased, secondary structures exert more and more influence on the stream dynamics, so that the magnitude of the corresponding spectral component increases concurrently.

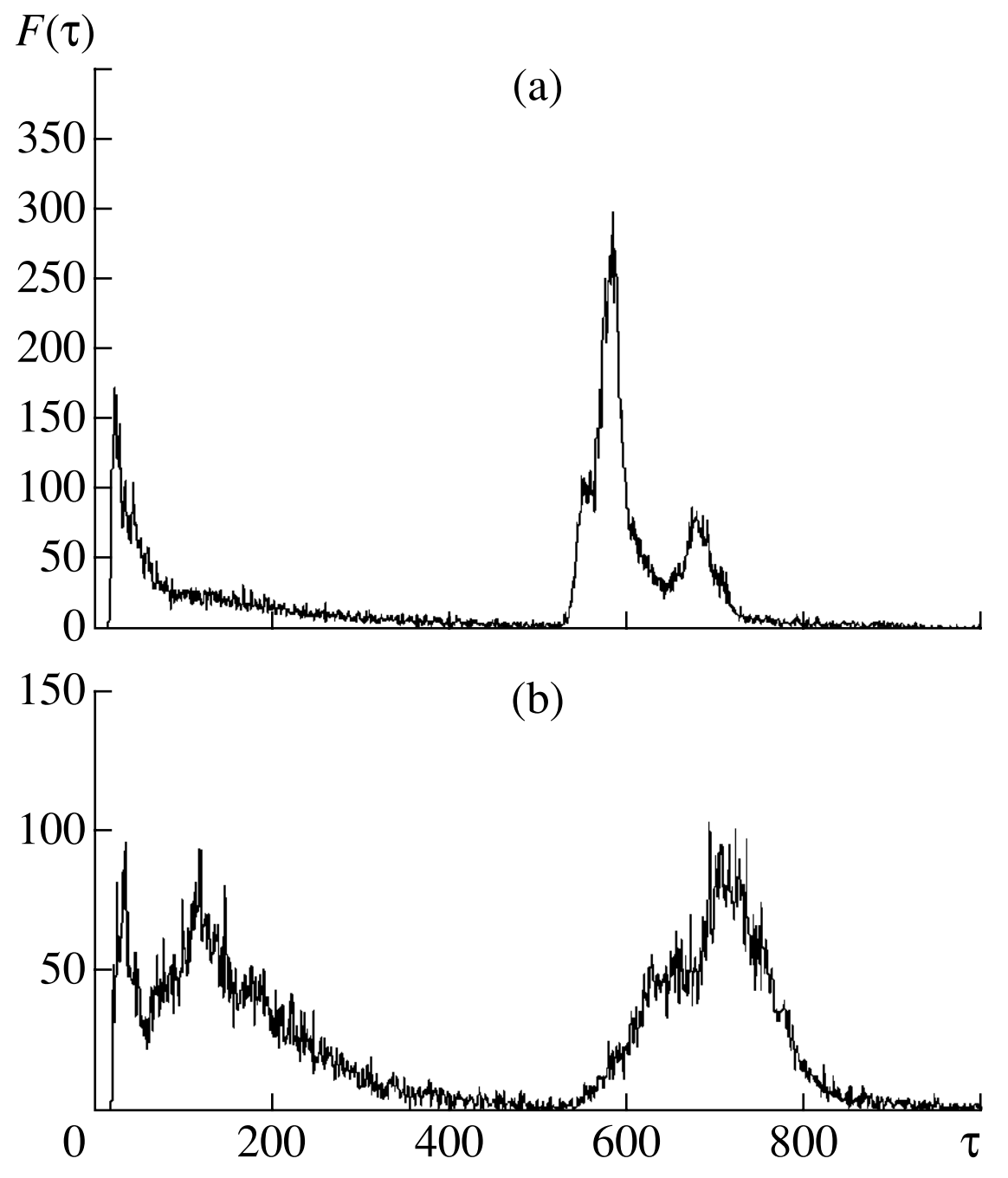

**Fig. 4.** Lifetime distributions of charged particles for regions (a) *A* and (b) *C* in Fig. 1.

## CONCLUSIONS

We have simulated the operation of a vircator klystron provided with an external feedback. It was found that the oscillator allows one to efficiently handle the formation of structures in the electron stream. If the input resonator is tuned to the natural frequency of the VC, three oscillation conditions are possible, depending on the feedback delay: (1) almost sinusoidal oscillation, (2) weak quasi-periodicity with two base frequencies (one of which is dominant), and (3) chaos in the form of fully developed two-periodic oscillation with two incommensurable frequencies. The onset of chaos is related to the formation of secondary structures in the electron stream as a result of its modulation, with the feedback delay being sufficiently large.

## ACKNOWLEDGMENTS

This study was supported by the Russian Foundation for Basic Research (project no. 99-02-16016).